\documentclass[sigplan,nonacm,screen]{acmart}
\usepackage{fancyhdr}
\AtBeginDocument{%
    \addtolength{\footskip}{2.0\baselineskip}%
    \fancyfoot[L]{\textit{\textbf{Author's preprint version.}}}%
}

\usepackage{Burst}

\begin{document}

\title{FaaS Is Not Enough: Serverless Handling of Burst-Parallel Jobs}


\author{Daniel Barcelona-Pons}
\email{daniel.barcelona@urv.cat}
\orcid{0000-0002-6051-9424}
\affiliation{%
  \institution{Universitat Rovira i Virgili}
  \city{Tarragona}
  \country{Spain}
}

\author{Aitor Arjona}
\email{aitor.arjona@urv.cat}
\orcid{0000-0001-5451-4865}
\affiliation{%
  \institution{Universitat Rovira i Virgili}
  \city{Tarragona}
  \country{Spain}
}

\author{Pedro García-López}
\email{pedro.garcia@urv.cat}
\orcid{0000-0002-9848-1492}
\affiliation{%
  \institution{Universitat Rovira i Virgili}
  \city{Tarragona}
  \country{Spain}
}

\author{Enrique Molina-Giménez}
\email{enrique.molina@urv.cat}
\orcid{0009-0005-2597-3815}
\affiliation{%
  \institution{Universitat Rovira i Virgili}
  \city{Tarragona}
  \country{Spain}
}

\author{Stepan Klymonchuk}
\email{stepan.klymonchuk@urv.cat}
\orcid{0009-0002-2360-1231}
\affiliation{%
  \institution{Universitat Rovira i Virgili}
  \city{Tarragona}
  \country{Spain}
}



\begin{abstract}



Function-as-a-Service (FaaS) struggles with burst-parallel jobs due to needing multiple independent invocations to start a job. The lack of a group invocation primitive complicates application development and overlooks crucial aspects like locality and worker communication.

We introduce a new serverless solution designed specifically for burst-parallel jobs. Unlike FaaS, our solution ensures job-level isolation using a group invocation primitive, allowing large groups of workers to be launched simultaneously. This method optimizes resource allocation by consolidating workers into fewer containers, speeding up their initialization and enhancing locality. Enhanced locality drastically reduces remote communication compared to FaaS, and combined with simultaneity, it enables workers to communicate synchronously via message passing and group collectives. This makes applications that are impractical with FaaS feasible. We implemented our solution on OpenWhisk, providing a communication middleware that efficiently uses locality with zero-copy messaging. Evaluations show that it reduces job invocation and communication latency, resulting in a $2\times$ speed-up for TeraSort and a $98.5\%$ reduction in remote communication for PageRank ($13\times$ speed-up) compared to traditional FaaS.

\end{abstract}






\maketitle


\begin{table}[t]
    \centering
    \footnotesize
    \caption{
        Start-up time of different cluster technologies compared to a FaaS service.
        AWS EMR Spark and GCP Dataproc use m5 and E2-standard VM families, respectively.
        Dask and Ray are deployed on managed m6i family EC2 VMs.
    }
    \label{Tab:cluster_startup}
    \begin{tabular}{lccc}
        \toprule
        Technology & Total vCPUs & Nodes & Start-up time  \\
        \midrule
        \multirow{2}{*}{EMR Spark} & \multirow{2}{*}{96}  &    6 & \second{296}  \\
        \null                      & \null                &   24 & \second{431}  \\
        \multirow{2}{*}{Dataproc}  & \multirow{2}{*}{96}  &    6 & \second{ 95}  \\
        \null                      & \null                &   24 & \second{113}  \\
        \multirow{2}{*}{Dask}      & \multirow{2}{*}{128} &    8 & \second{184}  \\
        \null                      & \null                &   64 & \second{253}  \\
        \multirow{2}{*}{Ray}       & \multirow{2}{*}{128} &    8 & \second{187}  \\
        \null                      & \null                &   64 & \second{229}  \\
        AWS $\lambda$ \gbyte{10}   & 6000                 & 1000 & \second{  6}  \\
        \bottomrule
    \end{tabular}
\end{table}

\section{Introduction}
\label{Introduction}

Managing a data processing platform to support varying volumes of data, in unpredictable patterns, and expecting real-time or interactive responses is an arduous task.
Current technologies like Spark or Dask face limitations in handling these cases due to their cluster-centric design, which prioritizes cluster utilization but suffers from critical resource misprovisioning~\cite{paris_serverless_2020}.
Their elasticity mechanisms are too slow to match changes in demand, so clusters either stay underprovisioned and cannot fulfill real-time processing objectives, or they are largely overprovisioned for peak demand and become very expensive but mostly idle~\cite{tariq_sequoia_2020, kaffes_centralized_2019}.

Function-as-a-Service (FaaS) has gained traction as a solution to the resource provisioning problem as it offers rapid, on-demand, no-ops scaling and a pay-as-you-go billing model at very fine granularity (MB per ms).
Its resource \emph{burstability} has set FaaS aside from traditional cluster technologies (see \cref{Tab:cluster_startup}) and several research works~\cite{jonas_occupy_2017, fouladi_encoding_2017, sampe_dataanalytics_2018, fouladi_gg_2019} have used thousands of short-lived functions working in parallel in data- and compute-intensive tasks such as data analytics, video encoding, or code compilation.

This has brought a new concept in cloud computing that refers to the ability to quickly respond to sudden, parallel workloads without provisioning a cluster in advance.
E.g., \citet{fouladi_encoding_2017} talk about a ``burstable supercomputer-on-demand'' and a ``burst-parallel swarm of thousands of cloud functions, all working on the same job.''
More recently, the concept of flash burst~\cite{li_millisort_2021} appeared for applications that use a large number of servers for a very short time (down to a millisecond).
However, they all coincide that the current model supported by FaaS is too narrow and precluding for massively parallel data processing programs (MPP)~\cite{berkeley_serverless_2019}.

In this work, we highlight that the key issue of FaaS hindering burst-parallel jobs is its lack of group awareness.
Indeed, FaaS users need multiple independent service calls to spawn a fleet of workers, which become strongly isolated from each other.
We note that such fine-grained isolation is damaging and unnecessary for collaborative jobs, and thus propose to raise the multi-tenant boundaries to the job level.

We present \emph{burst computing}, a new cloud computing model to deal with quick, sudden, massively parallel workloads, which we call \emph{bursts}.
To this end, the service offers a \emph{group invocation} primitive to handle the whole job as a unit.
This allows to optimize resource allocation, ensure worker parallelism, and perform \emph{packing}: running multiple workers co-located in the same environment.
Besides the speed-up in worker start-up latency, this enables \emph{worker locality}, which can be exploited to improve code and data loading effectively, and to aid powerful worker-to-worker communication patterns (e.g., broadcast, all-to-all) that seamlessly leverage shared memory channels with zero-copy mechanisms.

Burst computing is versatile to many applications, some currently unfeasible in FaaS.
Examples are data analytics and machine learning (e.g., TeraSort, TPC-DS, $k$-means), stream processing (e.g., video encoding or analytics), or other compute-intensive workloads (e.g., code compilation or simulations).
Bursts may be stateless (grid search, Monte Carlo simulations) or stateful (table joins, aggregations).


We make the following contributions:
\begin{compactitem}
    \item We present burst computing, a novel cloud computing model for short, sudden, massively parallel jobs (bursts).
    To the best of our knowledge, no cloud vendor or research effort has created the necessary infrastructure to support these workloads.
    \item Burst computing evolves FaaS with a key novel group invocation primitive (a flare) that raises multi-tenant isolation from a single function invocation to the whole job.
    In consequence, the system launches massive process groups with guaranteed parallelism and packs workers together to exploit locality.
    \item We implement a burst computing platform by extending OpenWhisk, a state-of-the-art FaaS system.
    Our implementation includes a specialized Rust worker runtime and a burst communication middleware that seamlessly leverage worker locality with collective code/data loading and zero-copy memory messaging.
    \item We evaluate our platform on several burst-parallel workloads against a FaaS solution.
    Burst computing improves job invocation latency (up to $11.5\times$ faster), worker simultaneity (up to $26.5\times$ lower MAD), and group communication (up to $98\%$ in a broadcast), for a speed-up of $13\times$ in PageRank and $2\times$ in TeraSort.
\end{compactitem}


\section{Motivation: in search of burstability}
\label{Motivation}

\begin{figure}[t]
  \centering
  
  \includegraphics[]{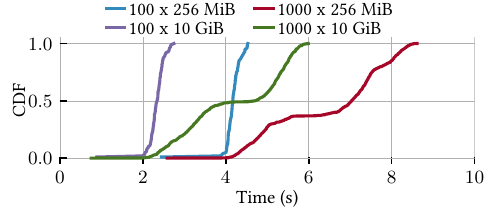}  

  \caption{
    CDF of FaaS function start-up time (cold start) in AWS Lambda for 100 and 1000 function invocations on two memory configurations.
  }
  \label{Fig:faas_startup}
\end{figure}

Many works are turning toward serverless services to perform massive data processing jobs~\cite{fouladi_encoding_2017,fouladi_gg_2019,jonas_occupy_2017,ao_sprocket_2018,werner_referenceSDP_2024,carver_wukong_2020} even though the current offer (mainly FaaS) is not designed for such workloads~\cite{berkeley_serverless_2019,hellerstein_serverless_2018,barcelonapons_benchmarking_2020}.
Reasons are varied, but a central point is the resource \emph{burstiness} that serverless offerings usually provide~\cite{muller_serverless_2020}.
Through serverless burstiness, many applications benefit from on-demand resources at very fine granularity and pay very precisely only for what they need, when they need it.
Further, they access those resources free of management (no-ops) at a virtually infinite scale.
In some cases, immediate access to unlimited resources is a need.
Think of, for instance, a scientist-driven interactive workflow doing different analyses on huge data sets, dynamically changing parameters that heavily affect the workload volume.
Or real-time application data streams or video feeds that must be critically processed online while its volume and analytics complexity may vary extremely over time.

We use the term \emph{burst} to refer to these types of serverless applications because they happen in bursts of computation.
In sum, bursts are massively parallel processing (MPP) workloads that appear suddenly and process large and variable amounts of data in a very short time (1 or 2 minutes or less).

Current data processing solutions such as Spark, Dask, Flink, or Ray start to fail to support bursts.
Even by leveraging the cloud's elasticity, their reaction to workload changes is too slow.
The key reason is that they are cluster-centric and actively manage capacity at coarse granularity (VMs) with high start-up time and operational complexity~\cite{muller_serverless_2020}.
\cref{Tab:cluster_startup} shows that starting one of these technologies is intolerable for critical real-time applications.
Even a fully cloud-managed version of Spark (``serverless Spark'') takes minutes to start just a few nodes~\cite{aws_bestEMR_2018}, clearly inappropriate for running short jobs interactively or at real time.

Contrarily, FaaS services provide a large scale compute substrate much faster.
\cref{Fig:faas_startup} shows that AWS Lambda may spawn a fleet of 100 functions in less than \second{4}, or 1000 in \second{6} (cold start); a much more appropriate time range for bursts.\footnote{
Note that, interestingly, \cref{Fig:faas_startup} shows a bigger invocation delay for small functions \mbyte{256} than large ones (\gbyte{10}).
This also happens on other providers (e.g., GCP) and it can be explained by the cost or scheduling complexity of finding and isolating finer resources.
}
To wit, serverless provides resource \emph{burstability} that allows to move from a cluster-centric view to use what resources are available, to a new job-centric view to use precisely what the job needs from the cloud pool of infinite resources.

\subsection{FaaS is holding us back}
\label{Motivation:Problem}

A review of literature will show us that running MPP programs (and thus bursts) atop FaaS brings many challenges~\cite{hellerstein_serverless_2018,berkeley_serverless_2019,muller_serverless_2020}.
We highlight three friction points:
\begin{inparaenum}
    \item[(\textbf{F1})] worker isolation, 
    \item[(\textbf{F2})] job fragmentation with complex orchestration, and
    \item[(\textbf{F3})] huge data movement.
\end{inparaenum}

\begin{figure}[t]
  \centering
  \includegraphics[width=\linewidth]{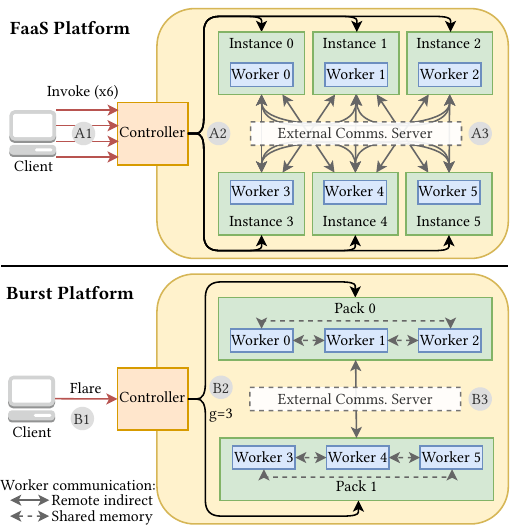}
  \caption{
    Running a data processing job of 6 workers in FaaS and burst computing with granularity 3.
  }
  \label{Fig:burst2}
\end{figure}

To illustrate these points, let's follow the execution of a parallel job atop a FaaS platform in \cref{Fig:burst2}.
Suppose we want to run a parallel job with 6 workers.
The job could be embarrassingly parallel (stateless) such as hyperparameter tuning, or require the workers to synchronize and share state at some point (stateful) such as TeraSort or, more intensively due to its iterative nature, PageRank.

Friction \textbf{F1} appears because multi-tenant isolation is at the level of a function invocation.
FaaS only allows to spawn function instances independently, one at a time, so we must make multiple service HTTP requests to obtain the 6 workers (\textbf{A1}).
Besides the added latency of repetitive requests, this is an issue for parallel jobs because the platform is not aware of these workers being collaborators, and thus cannot guarantee their parallelism.
This creates delays or skews between workers that potentially harm job execution.
Take for instance the 1000 invocations back in \cref{Fig:faas_startup}, where the last function starts up to \second{6} after the first one.\footnote{
    Further evaluation (not shown in the plot) reveals that this delay may increase to \second{44} in GCP, or \second{20} in an on-premises OpenWhisk deployment.
} 
Even more, this lack of \emph{group awareness} in the platform forces to populate identical environments (instances) for each invocation (\textbf{A2}), which stresses the system with code, dependency, and data\footnote{
For instance, hyperparameter tuning uses the same data in all workers.
} loading that create memory duplication~\cite{qiu_deduplication_2023,stojkovic_mxfaas_2023}.

\begin{figure}[t]
  \centering
  \includegraphics[width=\linewidth]{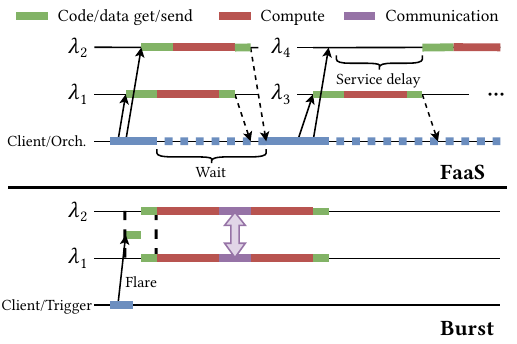}
  \caption{
    Timeline of a parallel job in FaaS and burst computing approaches.
  }
  \label{Fig:burst}
\end{figure}

Friction \textbf{F2} occurs in cases where workers need to coordinate at some point.
For instance, TeraSort à la MapReduce includes a data shuffle amidst the job, and PageRank iteratively aggregates a vector.
Because simultaneity is not guaranteed, workers may not exist at the same time and thus cannot communicate.
Instead, we must split the job into several stages where workers read and write intermediate data through an external storage solution (asynchronously).
Such pattern is depicted in \cref{Fig:burst}, it creates job fragmentation, and complicates its orchestration.
Iterative algorithms like PageRank or $k$-means that constantly aggregate data are unfeasible with this approach~\cite{barcelonapons_crucial_2019}.
On the one hand, it increases data movement and requires worker recreation at each stage (adding code and data loading overhead).
On the other hand, it necessitates an active orchestration process that lives throughout the job (mostly idle) to monitor the state of workers and oversee the overall job progress.\footnote{
This can be painful since FaaS does not provide monitoring mechanisms.
}

These issues are emphasized by friction \textbf{F3}.
Because we have many tiny isolated workers, most communication patterns (e.g., a shuffle) require numerous remote connections (\textbf{A3}), which in data processing workloads may result in very large data transfers.
Since functions cannot communicate directly, indirect communication worsens the trouble~\cite{lu_serialization_2024,copik_fmi_2023}.

\section{Burst Computing}
\label{Burst}

Burst computing is a novel paradigm to run \emph{bursts} in the cloud.
It overcomes the above frictions with two key principles that evolve FaaS: group awareness and locality exploitation.
\cref{Fig:burst2} shows these differences.

The key problem to run massively parallel jobs in FaaS is that multi-tenant isolation is at the level of a single function invocation (\textbf{F1}) and workers are precluded from collaboration.
Because a job belongs to a single tenant, it makes sense to raise the level of isolation to the job and handle all its workers as a group.
To this end, burst computing provides a group invocation primitive, which we call \emph{flare} (\textbf{B1}), to instantly launch massive process groups with guaranteed parallelism.
To the best of our knowledge, we are the first to implement this kind of primitive in a serverless system.
Flares bring group awareness to the platform, which is key to perform worker \emph{packing} (\textbf{B2}).
In particular, the platform spawns several job workers in the same environment to optimize resource allocation.
In our sample job with 6 workers (we say the \emph{burst size} is 6) in \cref{Fig:burst2}, if we set up a \emph{pack granularity} of 3 ($g=3$), the platform would only spawn 2 packs, each with 3 workers.
Packing establishes \emph{worker locality} and allows to optimize environment generation with less container creations, and collective code and data loading.

We avoid friction \textbf{F2} because workers may communicate synchronously due to their guaranteed parallelism and because the platform gives them job context (e.g., burst size or worker IDs).
This allows communication patterns previously unfeasible in FaaS, such as worker-to-worker message passing and collectives, that simplify job orchestration.
This difference is depicted in \cref{Fig:burst}.
Friction \textbf{F3} is addressed because communication (\textbf{B3}) can seamlessly exploit locality and use shared memory communication between workers in the same pack, which heavily reduces remote transfers.

We put these ideas into a prototype burst computing platform design that we describe in \cref{Architecture}.
But first, we dive deeper into the details that allow our platform to exploit group awareness and worker locality from two points of view: the platform and the application.

\paragraph{Worker packing}
To exploit locality, the burst platform allocates workers into \emph{packs}.
A pack of workers run in the same container or environment.
The number of workers per pack is the burst's \emph{granularity}.
The higher the granularity, the lower the number of environments (packs) are created, resulting in a direct advantage in allocation time, which is a critical part of function invocation in FaaS.
Then, worker code and dependencies are loaded only once per pack and shared by all co-located workers.
This further helps with initialization time, significantly when dependencies are large, and optimizes resource usage (e.g., avoiding memory duplication~\cite{qiu_deduplication_2023}).
A similar reasoning applies for data loading:
workers processing the same data (like in hyperparameter tuning) download it just once per pack and utilize their aggregated resources to speed up the transfer (i.e., parallel downloads).

Choosing the burst granularity is a trade-off between ease of system management and locality maximization.
To illustrate that, we identify three strategies for packing workers in burst computing:
\begin{inparaenum}[(i)]
    \item heterogeneous packing, where workers are placed in containers as big as possible in the underlying system machines;
    \item homogeneous packing, where workers are placed in fixed-size containers (e.g., packs with 6 vCPUs%
    like the biggest AWS Lambda configuration); and
    \item mixed packing, where workers are put in fixed-size packs, but if multiple packs fall onto the same machine, they are merged into a single container.
\end{inparaenum}
The first approach maximizes locality, but it can become a resource scheduling problem as it is prone to fragmentation.
The homogeneous packing mitigates that issue, but it restricts worker locality.
The third strategy is the compromise that allows a fast and flexible management while still maximizing locality (see \cref{Evaluation}).

\paragraph{Worker communication}
To exploit locality, the burst application is \emph{elastically distributed and collaborative}.
First, applications are aware of being distributed in a group of workers and accept any worker multiplicity transparently.
Then, because workers are guaranteed to be parallel, they may coordinate synchronously.
This allows to perform common communication patterns between workers that simplify job orchestration in a single stage, but most importantly the connections seamlessly exploit worker locality, and thus greatly reduce remote data transfers. 

To this end, we design a burst worker-to-worker communication middleware that adopts well-known abstractions (e.g., send/receive, broadcast, or all-to-all) and heavily benefits from worker locality.
Our middleware seamlessly identifies messages between workers placed in the same pack and transmits them through zero-copy memory sharing.
Only messages between packs are transferred remotely, and the middleware optimizes these connections (e.g., a broadcast only sends one message per pack).
Connections between packs (remote) may be direct or indirect, and our middleware may implement different remote backends.
Our contributions are independent of this choice because burst computing reduces any remote communication through exploiting locality.
In this work, we follow the usual approach in FaaS and only consider indirect solutions using an external communication server (\textbf{B3}).
\cref{Evaluation} evaluates several backends.

\section{Design and implementation}
\label{Architecture}

\begin{figure}[t]
  \centering
  \includegraphics[width=\linewidth]{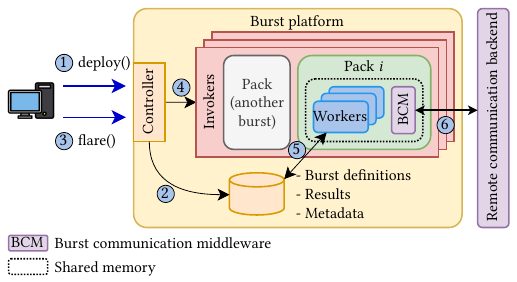}
  \caption{
    Burst computing platform overview.
  }
  \label{Fig:burst-arch}
\end{figure}

This section provides the design and implementation details of our prototype burst computing platform and communication middleware.
\cref{Fig:burst-arch} presents the overview of the main platform components and their interactions.

The burst platform extends the design of a FaaS platform to implement group invocation and worker packing.
We use Apache OpenWhisk as a basis and our platform shares its components.
We detail our modifications in \cref{Architecture:Platform}.
User interaction with the platform is managed by the \emph{controller}, which handles inbound HTTP requests to deploy and invoke bursts, oversees system resources, and performs worker packing.
The platform includes a \emph{database} to store burst definitions and configuration, as well as results and execution metadata.
Compute resources in the platform are provided by the \emph{invokers}, a set of machines with capacity for burst \emph{packs}.
Packs are run in containers that isolate a custom runtime environment that manages a set of workers.

Our burst communication middleware (BCM) has two main components: the core communication library and the remote backends.
The \emph{library} allows workers to send messages to each other or perform collectives, and it seamlessly leverages zero-copy or remote data transmission based on worker locality.
The library is extensible with multiple \emph{backends} that utilize different technologies to deliver messages remotely (e.g., Redis, RabbitMQ, or S3).

\subsection{Life cycle overview}
\label{Architecture:Lifecycle}


\cref{Fig:burst-arch} shows the life cycle of interaction with a burst computing platform to deploy and invoke a burst.
To deploy a new burst definition, the user must first \circled{1} contact the platform with a \texttt{deploy} HTTP request.
The controller receives the request and \circled{2} registers the burst definition into the database.
Later, when the user desires to trigger the execution of the burst, they \circled{3} send a \texttt{flare} HTTP request with specific parameters.
The controller handles the invocation and \circled{4} decides worker allocation based on the current state of the invoker machines.
The affected invokers receive the task to spawn the required runtime environments (packs) with space for as many workers as needed.
When the environments boot, their host invoker tells them which burst definition and parameters they should \circled{5} load from the database.
Then, each pack spawns its workers internally, which will execute the user-defined function (\texttt{work} in \cref{Tab:Inteface}) in parallel.
Workers may use the BCM to coordinate and share data.
This seamlessly uses shared memory or remote connections to \circled{6} communicate workers in the same or a different pack, respectively.
Additionally, workers may read or write data to external storage systems (e.g., object storage) or produce a result that is stored back to the database, where it may be retrieved later by users through another HTTP request.

\subsection{Developing and running bursts}
\label{Architecture:API}

User experience is key for burst computing.
As a serverless service, all resource management remains hidden.
Users interact with the service through a simple interface that allows to define bursts with resource-agnostic code and to schedule their execution.
This is similar to FaaS services that allow users to upload their function definitions and then set up triggers or invoke them as needed.
This section describes how users code and run bursts in a burst computing service, with the abstractions summarized in \cref{Tab:Inteface}.

\begin{table}[t]
  \centering
  \caption{
    Burst computing abstractions and API.
  }
  \label{Tab:Inteface}
  \begin{tabular}{ll}
    \toprule
    \textbf{Interface} & \textbf{Functions}  \\
    \midrule
    Burst           & \textbf{deploy}(\textit{defName}, \textit{package}, \textit{conf})           \\
    Service         & \hspace{0.5cm} upload and deploy a burst definition                               \\
    \null           & \textbf{flare}(\textit{defName}, [\textit{inputParams}])                        \\
    \null           & \hspace{0.5cm} invokes a burst                     \\
    [1ex]
    Burst           & abstract \textbf{work}(\textit{inputParams}, \textit{burstContext})                      \\
    Function        & \hspace{0.5cm} function to run on each worker                                     \\
    [1ex]
    Comm.           & \textbf{send}(\textit{data}, \textit{dest}) $\rightarrow$ none                    \\
    Primitives      & \textbf{recv}(\textit{source}) $\rightarrow$ data                                 \\
    \null           & \textbf{broadcast}(\textit{data}, \textit{root}) $\rightarrow$ data               \\
    \null           & \textbf{allToAll}(\textit{[data]}) $\rightarrow$ [data]                           \\
    \null           & \textbf{reduce}(\textit{data}, $f$(\textit{data}, \textit{data}) $\rightarrow$ \textit{data}) $\rightarrow$ data                                 \\
    \bottomrule
  \end{tabular}
\end{table}

\paragraph{Deployment}
Similar to FaaS functions, 
developers package and upload their burst definitions (code) to the cloud, giving them a name and configuration.
The configuration includes runtime parameters and worker characteristics (such as language and memory size), and burst-related parameters such as granularity preferences and application information.

\paragraph{Invocation}
Burst definitions are triggered for execution like functions in FaaS: an event or HTTP request notifies the intent to execute a burst with specific input parameters.
We call each burst invocation a \emph{burst flare} (\cref{Tab:Inteface}).
The main difference with FaaS is that a flare will spawn a group of parallel workers (instead of a single function instance).
The service ensures that all workers run simultaneously and applies packing.
In our prototype, the burst size is explicit on the size of the \textit{inputParams} array.\footnote{Smart burst sizing is left for future work. E.g., the platform may calculate the number of workers based on the input data volume (data-driven).}

\paragraph{Coding}
Burst definitions are coded as a single function which is run by each worker in the burst (\texttt{work} in \cref{Tab:Inteface}).
This function must be programmed elastically so that it accepts and runs correctly for any burst size.
The code is also agnostic of the packing performed by the service.
To that end, the \texttt{work} function receives a burst context object through which each worker may obtain its unique identifier (ID) within the particular flare and the burst size to adapt its logic to a specific portion of the problem and data.
This context object also gives access to the BCM.

\paragraph{Communication interface}
The BCM offers simple yet powerful worker-to-worker communication through message passing similar to MPI.
The abstractions are elastic (adapt to the burst size) and available through the burst context.
Burst computing programs make use of two basic primitives to connect workers: send and receive.
These primitives enable point to point communication between workers and are designed to send arbitrary volumes of data efficiently within the burst.
To facilitate common communication patterns in parallel jobs, bursts may also use group collectives.
As listed in \cref{Tab:Inteface}, our prototype implements broadcast, all-to-all, and reduce.
Primitives and collectives are locality-aware, although the programs remain agnostic to it, i.e., co-located workers (same pack) communicate on shared memory and only remote workers hit the network.

\begin{algorithm}[t]
\caption{
    Simplified source code of the PageRank \textit{work} function for burst computing.
    The accesses to the burst context to obtain the worker ID or communicate are highlighted.
}
\label{Alg:app}
\footnotesize

\begin{Verbatim}[commandchars=\\\{\}]
\PY{k}{fn} \PY{n+nf}{work}\PY{p}{(}\PY{n}{params}: \PY{n+nc}{Input}\PY{p}{,}\PY{+w}{ }\PY{n}{burst}: \PY{k+kp}{\PYZam{}}\PY{n+nc}{BurstContext}\PY{p}{)}\PY{+w}{ }\PYZhy{}\PYZgt{} \PY{n+nc}{Output}\PY{+w}{ }\PY{p}{\PYZob{}}
\PY{+w}{  }\PY{k+kd}{let}\PY{+w}{ }\PY{n}{num\PYZus{}nodes}\PY{+w}{ }\PY{o}{=}\PY{+w}{ }\PY{n}{params}\PY{p}{.}\PY{n}{num\PYZus{}nodes}\PY{p}{;}
\PY{+w}{  }\PY{k+kd}{let}\PY{+w}{ }\PY{k}{mut}\PY{+w}{ }\PY{n}{page\PYZus{}ranks}\PY{+w}{ }\PY{o}{=}\PY{+w}{ }\PY{n+nf+fm}{vec!}\PY{p}{[}\PY{l+m+mf}{1.0}\PY{+w}{ }\PY{o}{/}\PY{+w}{ }\PY{n}{num\PYZus{}nodes}\PY{p}{;}\PY{+w}{ }\PY{n}{num\PYZus{}nodes}\PY{p}{]}\PY{p}{;}
\PY{+w}{  }\PY{k+kd}{let}\PY{+w}{ }\PY{k}{mut}\PY{+w}{ }\PY{n}{sum}\PY{+w}{ }\PY{o}{=}\PY{+w}{ }\PY{n+nf+fm}{vec!}\PY{p}{[}\PY{l+m+mf}{0.0}\PY{p}{;}\PY{+w}{ }\PY{n}{num\PYZus{}nodes}\PY{p}{]}\PY{p}{;}
\PY{+w}{  }\PY{k+kd}{let}\PY{+w}{ }\PY{n}{adjacency\PYZus{}matrix}\PY{+w}{ }\PY{o}{=}\PY{+w}{ }\PY{nb}{get\PYZus{}adjacency\PYZus{}matrix}\PY{p}{(}\PY{o}{\PYZam{}}\PY{n}{params}\PY{p}{)}\PY{p}{;}
\PY{+w}{  }\PY{k}{while}\PY{+w}{ }\PY{n}{err}\PY{+w}{ }\PY{o}{\PYZlt{}}\PY{+w}{ }\PY{n}{ERROR\PYZus{}THRESHOLD}\PY{+w}{ }\PY{p}{\PYZob{}}
\PY{+w}{    }\PY{n}{page\PYZus{}ranks}\PY{+w}{ }\PY{o}{=}\PY{+w}{ }\PY{s+cs}{burst}\PY{s+cs}{.}\PY{s+cs}{broadcast}\PY{p}{(}\PY{n}{page\PYZus{}ranks}\PY{p}{,}\PY{+w}{ }\PY{n}{ROOT\PYZus{}WORKER}\PY{p}{)}\PY{p}{;}
\PY{+w}{    }\PY{k}{for}\PY{+w}{ }\PY{p}{(}\PY{n}{node}\PY{p}{,}\PY{+w}{ }\PY{n}{links}\PY{p}{)}\PY{+w}{ }\PY{k}{in}\PY{+w}{ }\PY{n}{graph}\PY{+w}{ }\PY{p}{\PYZob{}}
\PY{+w}{      }\PY{k}{for}\PY{+w}{ }\PY{n}{link}\PY{+w}{ }\PY{k}{in}\PY{+w}{ }\PY{n}{links}\PY{+w}{ }\PY{p}{\PYZob{}}
\PY{+w}{        }\PY{n}{sum}\PY{p}{[}\PY{o}{*}\PY{n}{link}\PY{p}{]}\PY{+w}{ }\PY{o}{+}\PY{o}{=}\PY{+w}{ }\PY{n}{page\PYZus{}ranks}\PY{p}{[}\PY{o}{*}\PY{n}{node}\PY{p}{]}\PY{+w}{ }\PY{o}{/}\PY{+w}{ }\PY{nb}{out\PYZus{}links}\PY{p}{(}\PY{o}{*}\PY{n}{node}\PY{p}{)}\PY{p}{;}
\PY{+w}{      }\PY{p}{\PYZcb{}}
\PY{+w}{    }\PY{p}{\PYZcb{}}
\PY{+w}{    }\PY{k+kd}{let}\PY{+w}{ }\PY{n}{reduced\PYZus{}ranks}\PY{+w}{ }\PY{o}{=}\PY{+w}{ }\PY{s+cs}{burst}\PY{s+cs}{.}\PY{s+cs}{reduce}\PY{p}{(}\PY{n}{sum}\PY{p}{,}\PY{+w}{ }\PY{o}{|}\PY{n}{vec1}\PY{p}{,}\PY{+w}{ }\PY{n}{vec2}\PY{o}{|}\PY{+w}{ }\PY{p}{\PYZob{}}
\PY{+w}{      }\PY{n}{vec1}\PY{p}{.}\PY{nb}{zip}\PY{p}{(}\PY{n}{vec2}\PY{p}{)}\PY{p}{.}\PY{nb}{map}\PY{p}{(}\PY{o}{|}\PY{p}{(}\PY{n}{a}\PY{p}{,}\PY{+w}{ }\PY{n}{b}\PY{p}{)}\PY{o}{|}\PY{+w}{ }\PY{n}{a}\PY{+w}{ }\PY{o}{+}\PY{+w}{ }\PY{n}{b}\PY{p}{)}\PY{p}{.}\PY{nb}{collect}\PY{p}{(}\PY{p}{)}
\PY{+w}{    }\PY{p}{\PYZcb{}}\PY{p}{)}\PY{p}{;}
\PY{+w}{    }\PY{k}{if}\PY{+w}{ }\PY{s+cs}{burst}\PY{s+cs}{.}\PY{s+cs}{worker\PYZus{}id}\PY{+w}{ }\PY{o}{=}\PY{o}{=}\PY{+w}{ }\PY{n}{ROOT\PYZus{}WORKER}\PY{+w}{ }\PY{p}{\PYZob{}}
\PY{+w}{      }\PY{n}{err}\PY{+w}{ }\PY{o}{=}\PY{+w}{ }\PY{nb}{calculate\PYZus{}error}\PY{p}{(}\PY{o}{\PYZam{}}\PY{n}{page\PYZus{}ranks}\PY{p}{,}\PY{+w}{ }\PY{o}{\PYZam{}}\PY{n}{reduced\PYZus{}ranks}\PY{p}{)}\PY{p}{;}
\PY{+w}{      }\PY{n}{page\PYZus{}ranks}\PY{+w}{ }\PY{o}{=}\PY{+w}{ }\PY{n}{reduced\PYZus{}ranks}\PY{p}{;}
\PY{+w}{    }\PY{p}{\PYZcb{}}
\PY{+w}{    }\PY{n}{err}\PY{+w}{ }\PY{o}{=}\PY{+w}{ }\PY{s+cs}{burst}\PY{s+cs}{.}\PY{s+cs}{broadcast}\PY{p}{(}\PY{n}{err}\PY{p}{,}\PY{+w}{ }\PY{n}{ROOT\PYZus{}WORKER}\PY{p}{)}\PY{p}{;}
\PY{+w}{    }\PY{nb}{reset\PYZus{}sums}\PY{p}{(}\PY{o}{\PYZam{}}\PY{k}{mut}\PY{+w}{ }\PY{n}{sum}\PY{p}{)}\PY{p}{;}
\PY{+w}{  }\PY{p}{\PYZcb{}}
\PY{+w}{  }\PY{n}{Output}\PY{+w}{ }\PY{p}{\PYZob{}}\PY{+w}{ }\PY{n}{page\PYZus{}ranks}\PY{+w}{ }\PY{p}{\PYZcb{}}
\PY{p}{\PYZcb{}}
\end{Verbatim}

\end{algorithm}

\subsection{Application example}
\label{Architecture:Example}


\cref{Alg:app} shows an example in Rust code (simplified) of the \texttt{work} function that implements the PageRank application.
The algorithm consists of an iterative process where each worker holds a portion of the adjacency graph (relating links between web pages).
On each iteration, the new global ranks are computed in parallel, aggregated, and reduced in a tree structure, then broadcasted from the root worker to the rest of them.
The algorithm runs until it converges past a threshold or reaches a limit of iterations.

Similar to the MPI computing model, all workers execute the same code but perform different logic based on the worker ID (the \textit{rank} in MPI).
The example highlights the worker accesses to the \texttt{BurstContext} object to perform collectives and obtain information about the current flare.
For instance, it is used to perform a collective \textit{broadcast} to share the updated ranks vector, and later, a \textit{reduce} to aggregate the partial ranks computed among the workers.
It also shows how a worker checks its ID when it needs to calculate the convergence, since this is only done by the root worker after collecting the aggregated vector in the reduce.



\subsection{Burst platform implementation}
\label{Architecture:Platform}

The prototype implementation is built on top of the popular Apache OpenWhisk open-source platform (v1.0.0).
We used OpenWhisk as a basis because it is a well-known, production-tested FaaS implementation.
Our changes amount to approximately \ksloc{2} and are available online.\footnote{
Soon available.
}
The modifications affect the main platform components, including the controller, the invoker, and the runtime environment.

The controller now supports two new HTTP endpoints for bursts: \texttt{deploy} and \texttt{flare}.
It also implements the logic to handle them as described in \cref{Architecture:Lifecycle}.
This includes the packing strategy in the three flavors defined in \cref{Burst}: heterogeneous, homogeneous, and mixed.
Granularity can be configured.
In any case, the controller calculates the number and size of packs based on the specific burst size and available resources in the invokers.

The invokers run a new monitoring logic that can be adjusted to report their load to the controller based on CPU instead of RAM.
Our prototype is configured to assign 1 vCPU per worker because bursts tend to be compute-intensive jobs and we do not consider parallelism within a worker,\footnote{The burst size (number of workers) determines its parallelism.} but other configurations are possible if required.
Invokers also implement new logic to support the creation and execution of packs, spawning Docker containers of the appropriate size and telling the runtime the number of workers to run, and their IDs and context.

For the runtime, we adapted the official OpenWhisk Rust environment, but it is possible to support others.
The new logic allows to spawn multiple workers within it as requested by its host invoker.
In particular, the Rust runtime spawns one thread per worker to provide parallelism.
Finally, the runtime also includes our BCM built-in.

\subsection{BCM implementation}
\label{Architecture:Communication}

The burst communication middleware (BCM) is coded in Rust in about \ksloc{5}.
It is readily available for our custom Rust runtime, but other languages may also use it through bindings.
The prototype is available online.\footnote{
Soon available.
}
It enables the transmission of intra-pack (zero-copy) and inter-pack (via remote backend) messages.
For this, each worker has context (provided by the invoker) about the particular flare to which it belongs, including the flare, pack, and worker IDs, the burst size, and the distribution of packs (which worker belongs to which pack).
The BCM is instantiated by the runtime (once per pack) and made available to workers as a parameter (in the \texttt{work} function as shown in \cref{Tab:Inteface}).

For local communication, BCM uses in-memory queues to send and receive data between workers in the same pack.
In the Rust runtime, workers are threads and reside in the same memory space, so shared memory mechanisms are not necessary (e.g. \texttt{shm\_open} or \texttt{mmap}).
Instead, workers just pass memory pointers between them.
Thanks to Rust's memory safety guarantees, access to shared data is thread-safe.
For example, the root worker in a \textit{broadcast} sends a read-only memory pointer to its local workers, and they safely access the message concurrently.
To modify the data, a receiver may use mechanisms such as copy-on-write.

For remote communication, each pack has a shared connection pool to the remote backend, which allows each worker within the pack to send and receive messages concurrently, with the goal of maximizing the container's bandwidth.
This is especially useful in primitives like \textit{all-to-all}, where all workers must open channels with all the others.
For large messages, the data is split into smaller chunks that are sent and received concurrently.
This maximizes network utilization and allows readers to start receiving data from the first chunk, instead of waiting for the full message to be available at the backend.

The BCM is extensible, allowing the implementation of more remote backends.
Currently, we support Redis, DragonflyDB, RabbitMQ, and S3.
The backend interface for remote communication differentiates between sending direct messages (one-to-one) and broadcast messages (one-to-many) because it affects how many times a message is read from the backend server.
E.g., in RabbitMQ, one-to-one messages use direct brokers, while one-to-many use fan-out brokers.


To ensure that no messages are lost (\textit{at-least-once} delivery semantics), the BCM keeps a count of direct messages sent between each pair of workers, and for each collective operation.
The middleware handles duplicate and/or out-of-order messages.
For that, messages include a header with the source and destination worker, collective type, counter, and, if chunked, the number of chunks and chunk number.
For chunked messages received out-of-order, a memory region is reserved for the total payload where chunks are written to their respective offset as they come in.

\section{Evaluation}
\label{Evaluation}

The goal of our evaluation is to show how burst computing compares to current FaaS for the three friction points described in \cref{Motivation:Problem}.
We analyze the effects of worker packing and locality on the platform and our communication middleware (BCM).
All experiments run on Amazon Web Services (AWS) in the same region (us-east-1).
Evaluation code is available online.\footnote{
Soon available.
}

\subsection{Burst group invocation}
Group invocation is the key element against friction \textbf{F1}.
Here we evaluate how job-level isolation improves worker readiness time (invocation latency), ensures their simultaneity, and provides locality for collaborative code and data loading.

\paragraph{Setup}
The burst platform runs on an Amazon EKS cluster, with the control plane on a \texttt{t4i.xlarge} VM (4~vCPUs and 16~GB RAM), and the invokers on up to 20 \texttt{c7i.12xlarge} VMs (48~vCPUs and 96~GB RAM).
This gives us space to accommodate up to $960$ workers with 1 vCPU each.

\begin{figure}
  \centering
  \includegraphics[width=\linewidth]{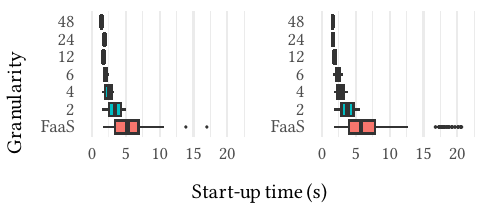}
  \caption{
    Burst start-up time for different packing granularity (worker latency distribution).
    Left and right show, respectively, burst sizes of 48 and 960.
  }
  \label{Fig:boxplot-invk}
\end{figure}

\paragraph{Impact on burst invocation latency}
First, we use the homogeneous packing policy to evaluate how assigning different granularity affects burst invocation latency.
The exploration is depicted in \cref{Fig:boxplot-invk}, showing bursts of sizes 48 and 960.\footnote{
We conducted experiments with similar results for burst sizes in-between.
}
It is quickly apparent that as the granularity increases, the start-up time decreases, and generally becomes more consistent across workers.
For instance, the latency of having all workers ready in a burst size of 960 reduces by $11.5\times$ from granularity 1 (FaaS) to 48.
We found that container creation dominates invocation latency, hence higher granularity performs best.
This proves that creating the biggest possible containers, and thus the less amount of them (heterogeneous packing), achieves the best start-up latency, since it creates a single container per invoker per flare.
By extension, the mixed packing strategy exhibits the same results, but allows the system to manage resources more effectively in small portions to facilitate allocation and avoid resource fragmentation.
To clearly evaluate the effects of granularity, the rest of the evaluation still uses homogeneous packing.

\begin{figure}
  \centering
  \begin{subfigure}[b]{\linewidth}
    \centering
    \includegraphics[width=0.8\linewidth]{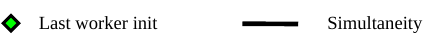}
    \captionsetup{labelformat=empty}
  \end{subfigure}
  \hfill
  \begin{subfigure}[b]{\linewidth}
    \centering
    \includegraphics[width=\linewidth]{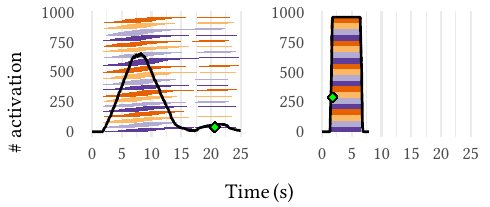}
    \captionsetup{labelformat=empty}
  \end{subfigure}
  \caption{Simultaneity in FaaS (left) and Burst with granularity 48 (right).
    Each horizontal bar represents the life-time of a worker.
    Colors denote invoker machines.
    }
  \label{Fig:invocation-simult}
\end{figure}

\paragraph{Impact on worker simultaneity}

To evaluate worker simultaneity, we run a burst job with size 960 on FaaS against burst computing with granularity 48.
For demonstration purposes, each worker performs a 5-second sleep and we plot their execution timeline in \cref{Fig:invocation-simult}.
The plot shows that burst computing achieves faster resource allocation and quicker readiness of workers.
This ensures worker parallelism.
Analyzing dispersity of worker start-up time (seen also in \cref{Fig:boxplot-invk}), the FaaS execution evinces a range of \second{18.8}, with a median average deviation (MAD) of \second{2.65}.
In contrast, the range in burst computing with granularity of 48 is just \second{0.44} (MAD is \second{0.1}).
Compared, the range is $43\times$ lower in burst computing, with MAD showing $26.5\times$ lower dispersity than FaaS.
Dispersity in worker start-up latency precludes FaaS to achieve full parallelism (all workers running simultaneously from start to finish), while burst guarantees it.

\begin{figure}
  \centering
  \includegraphics[width=\linewidth]{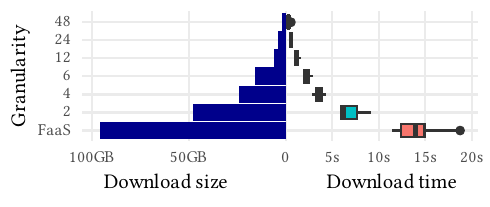}
  \caption{
    A burst of 96 workers loading the same \gbyte{1} object from S3 in different granularity.
  }
  \label{Fig:download-s3}
\end{figure}

\paragraph{Impact on data loading}
Burst computing mitigates the FaaS problem of loading the same data on all functions (see \cref{Motivation:Problem}).
Further, with burst computing we can use mechanisms to optimize this problem by downloading data once per pack (trivially reducing data ingestion), and using the parallelism within a pack to accelerate the transfer (leveraging parallel object storage byte range reads).
We evaluate this mechanism on multiple granularity setups and present it in \cref{Fig:download-s3}.
Burst optimizations achieve a download time speed-up of $32.6\times$ with granularity 48 compared to FaaS.

\paragraph{Takeaway}
Flares eliminate friction \textbf{F1} through faster worker group initialization ($11.5\times$) and ensured simultaneity ($43\times$ less dispersed workers) that enables locality with packing.
In turn, locality may accelerate data download in applications ($32.6\times$), tackling friction \textbf{F3}.

\subsection{Burst inter-pack communication}

Before we evaluate the effects of the BCM on frictions \textbf{F2} and \textbf{F3}, we want to ensure that an indirect communication model is feasible and to find a backend that sustains the load of bursts at scale. 
For this, we measure the throughput of several indirect communication backends.
Specifically, we test Redis, DragonflyDB (a Redis-compatible multi-threaded alternative), RabbitMQ, and S3.
Redis and DragonflyDB evaluate two flavors: using lists or streams.

\begin{figure}
    \centering
    \begin{subfigure}{\linewidth}
        \includegraphics[]{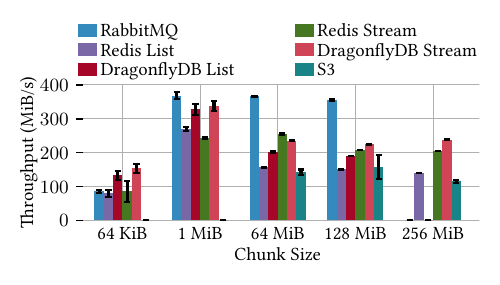}
        \caption{
        Throughput between two remote workers sending a \gbyte{1} payload chunked in different sizes.
        }
        \label{Fig:throughput-pair}
    \end{subfigure}
    \begin{subfigure}{\linewidth}
        \includegraphics[]{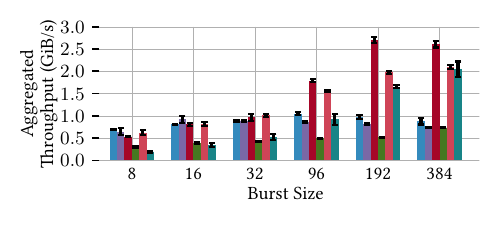}
        \caption{
        Aggregate throughput of two remote packs, A and B, of varying size, where each worker from pack A sends a \mbyte{256} payload to another worker from remote pack B.
        }
        \label{Fig:throughput}
    \end{subfigure}
    \caption{
        Throughput experiments for the RabbitMQ, Redis, DragonflyDB and S3 backends of the BCM.
        Median values with standard deviation (10 runs).
    }
    \label{Fig:sub-throughput}
\end{figure}

\paragraph{Message chunk size}
The BCM chunks messages into several blocks to optimize network utilization and allow parallel read/write. 
The optimal chunk size is a trade-off between latency to first byte and operation overhead, and it varies for each communication backend.
To find the optimal configuration, we measure the throughput of sending a \gbyte{1} message between two remote workers.
The workers run on two \texttt{c7i.large} machines (4 vCPUs, 8 GB) and we deploy a \texttt{c7i.16xlarge} (64 vCPUs, 128 GB) for the intermediate server.
\cref{Fig:throughput-pair} plots the results.
RabbitMQ offers a constant throughput for larger chunk sizes, but does not allow payloads larger than \mbyte{128} due to AMQP protocol limitations.
Redis and DragonflyDB work best at \mbyte{1}, the latter being slightly superior.
S3 offers the lowest throughput because object stores are not designed for small files (\mbyte{1} or less exceeds the allowed service request rate limits).

\paragraph{Maximum throughput}
To understand how the different backends scale under parallel load, we measure the aggregated throughput between several pairs of workers communicating simultaneously.
In this experiment, we launch a group of workers (burst size from 8 to 384) split into two remote groups.
Each worker in a group A sends a fixed message (\mbyte{256}) to a worker in the other, remote group B.
As the burst size increases, so does the total data volume sent.
Each backend uses the optimal chunk size assessed in the micro-benchmark above.
Workers run on two VMs scaled to the burst size (from \texttt{c7i.xlarge} for 8 workers to \texttt{c7i.48xlarge} for 384), and the communication server runs on one \texttt{c7i.48xlarge} instance.
The results are shown in \cref{Fig:throughput}.
We observe that RabbitMQ does not scale beyond \gbytepers{1}.
For the in-memory stores, the approach with lists performs better than streams.
Like RabbitMQ, Redis does not scale with parallelism because it is single-threaded.
In contrast, DragonflyDB does scale and achieves the highest throughput, surpassing \gbytepers{2.5} for large burst sizes.
S3 also scales with parallelism but remains slower.

\paragraph{Takeaway}
The BCM achieves high throughput even with indirect communication, and some backends sustain up to the evaluated 384 workers with individual connections, suggesting a feasible approach.
In view of the results, the rest of the evaluation uses DragonflyDB List with \mbyte{1} chunks.

\subsection{Burst group collectives}
\label{Sec:experiment_group_collectives}

This section shows the impact of locality on group collectives as a means to face friction \textbf{F3}.
To that end, we measure end-to-end latency, i.e., the total time it takes for all workers to complete the collective, as we vary the packing granularity.
We present results for the \textit{broadcast} and \textit{all-to-all} collectives.
\textit{Reduce} behaves similar to \textit{broadcast}, because they follow the same data movement patterns.\footnote{
Other collectives like \textit{gather} and \textit{scatter}, which we left for future work, are similar to \textit{all-to-all}.
}

\paragraph{Setup}
We employ one, two and four \texttt{c7i.12xlarge} VMs (48 vCPUs, 96 GB) for bursts of sizes 48, 96 and 192, respectively.
Granularity varies from 1 to 48.
Each worker uses \mbyte{256} as data for each collective call.
The backend server runs on one \texttt{c7i.48xlarge} VM (192 vCPUs, 384 GB).

\begin{figure}
    \centering
    \begin{subfigure}{\linewidth}
        \includegraphics[]{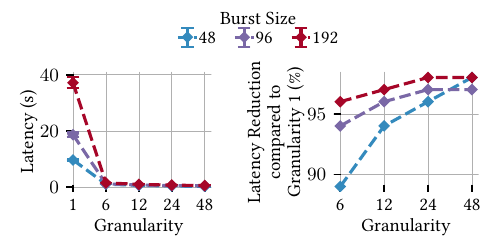}
        \caption{Broadcast}
        \label{Fig:broadcast}
    \end{subfigure}
    \begin{subfigure}{\linewidth}
        \includegraphics[]{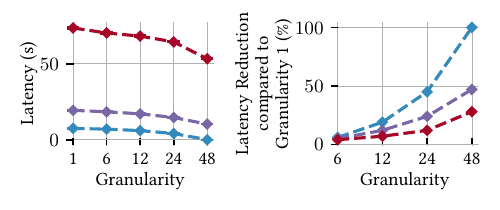}
        \caption{All-to-All}
        \label{Fig:all2all}
    \end{subfigure}
    \caption{
        Latency and latency reduction percentage with respect to granularity 1 of different collectives, for varying granularity and burst size.
    }
    \label{Fig:collectives}
\end{figure}

\paragraph{Results}
The results are shown in \cref{Fig:collectives}.
Overall, latency decreases as the granularity increases.
This is because remote communication is the main bottleneck of collective operations and it decreases in volume as granularity increases and more data movement becomes local.
The cost of local communication is irrelevant compared to the remote one.
Broadcast sends the message once, but reads it once per pack.
This means that remote data movement is directly proportional to the number of packs, i.e., if we halve the packs, we move half the data.
This becomes a fast decrease in latency as we increase pack granularity; near $98\%$ reduction with granularity 48 (\cref{Fig:broadcast}).
All-to-all is more intensive in data traffic because all workers have a message of \mbyte{256} for each of the other workers (\gbyte{48} total with 192 workers).
This means that even if we only have two packs, half the data traffic is still remote.
This is clearly evident in \cref{Fig:all2all}.
Considering the highest granularity (48), burst sizes 48, 96, and 192 create one, two, and four packs; thus the reduction in latency is approximately $100\%$, $50\%$, and $25\%$ respectively.

\paragraph{Takeaway}
Locality-aware group collectives heavily mitigate friction \textbf{F3} by seamlessly reducing remote data traffic.

\subsection{Burst applications}
We complete our evaluation with three real-world bursts.
These applications clearly show all friction points while providing an overall view of the effects of burst computing compared to FaaS-based implementations.

\subsubsection{Hyperparameter tuning}

Grid search is a machine learning technique in which a set of hyperparameter values are evaluated to find the combination that yields the best performance for a given model.
This evaluation is done in parallel, with each worker processing a full copy of the training dataset.
Since there are no dependencies between parallel tasks, FaaS seems suitable for this job.
However, each function would download a copy of the data, regardless of whether there are functions co-located on the same node.
This results in a waste of bandwidth and memory due to duplicate data downloads.
Burst computing provides an optimization opportunity by exploiting locality.

\paragraph{Setup}
The grid search is applied to a stochastic gradient descent model in a \texttt{sklearn} Python application.
We use a \mbyte{500} Amazon reviews dataset (CSV), available at Kaggle\footnote{\url{ https://www.kaggle.com/bittlingmayer/amazonreviews}} and stored in an S3 bucket.
AWS Lambda is the baseline (denoted as granularity 1), with a memory configuration of \mbyte{1769}, which provides a full vCPU.
The burst platform uses a \texttt{c7i.24xlarge} instance.

\begin{table}[t]
  \centering
  \footnotesize
  \caption{
    Time to start 96 workers and gather input data (ready for computation) in hyperparameter tuning for different burst granularity.
  }
  \label{Tab:Hypertuning}
  \begin{tabular}{l|cccccc}
  \toprule
     \textit{Granularity}    &  1 (FaaS) &  6    & 12   & 24   & 48   & 96    \\
     \textit{Ready time (s)} &     17.51 &  5.65 & 3.64 & 3.18 & 2.96 & 2.57  \\
  \bottomrule
  \end{tabular}
\end{table}

\paragraph{Results}
\cref{Tab:Hypertuning} collects the results.
``Ready time'' refers to the time elapsed from client-side job invocation until the input data is available to all workers and they are ready to compute.
We see how burst computing quickly reduces this time as we increase the granularity.
This effect has two causes.
First, the group invocation primitive speeds up invocation time compared to FaaS, from approximately 4 to \second{1.5} with granularity 96.
Second, data download can be optimized as assessed in \cref{Fig:download-s3}.
While FaaS has to download a copy of the data on each worker, workers co-located in a pack collaborate in downloading the input in parallel.
Hence, as we increase the granularity, the input download time decreases, going from \second{14} in FaaS to \second{1} with granularity 48 or 96.

\subsubsection{PageRank}

PageRank is a well-known data analytics workload with intensive worker coordination.
It involves iterative and heavy data aggregation for large datasets, which is of interest for benchmarking worker communication in burst computing.
We adapted PageRank for burst computing from the MapReduce version in the Hi-Bench suite~\cite{huang_hibench_2010}.
\cref{Architecture:Example} details the implementation.
We skip reporting the MapReduce version atop FaaS because the number of (short) stages necessary to perform the iterative aggregations make it obviously slower.

\paragraph{Setup}
This experiment uses four \texttt{c7i.16xlarge} VMs (64 vCPU, 128~GB).
The graph dataset is generated with Hi-Bench~\cite{huang_hibench_2010} consisting of \million{50} nodes (\gbyte{$\approx30$}) in 256 partitions.
The algorithm runs over 10 iterations, with a burst size of 256, varying the granularity between 1 and 64.

\begin{figure}
  \centering
  \includegraphics[]{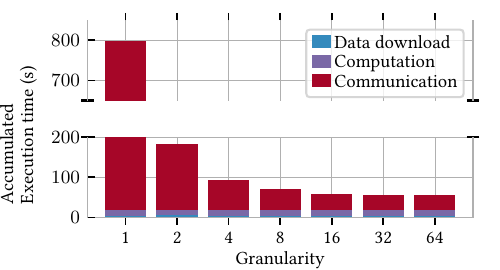}
  \caption{
  Average aggregated time of each phase in Page-Rank for different granularity.
  }
  \label{Fig:pagerank-exec-time}
\end{figure}

\paragraph{Results}
\cref{Fig:pagerank-exec-time} shows the total execution time of all iterations split into phases: the time dedicated to download input data (from S3), compute the ranking, and communicate (collectives) between workers.
Times for each phase are averaged across workers, and all iterations added.
\cref{Tab:pagerank-traffic} shows the network traffic and \% of reduction compared to granularity 1.
Communication accounts for the majority of the execution time because the rank vector must be aggregated and shared at each iteration.
Our configuration uses a vector of \mbyte{40} that is sent, received, and aggregated in a tree pattern across the workers, and then broadcast from the root worker to the rest of them.
As granularity increases, the remote portion of this movement decreases.
For example, with a granularity of 2, only the first level of the (binary) reduction tree is local (the leaves), and the rest communicate remotely.
With granularity at 64, there are 4 packs, so remote communication occurs only in the last 2 tree levels.
With this setup, we achieve a 98.5\% reduction in data traffic and a $13\times$ speed-up compared to granularity 1.

\begin{table}[t]
    \footnotesize
    \centering
    \caption{Aggregated network traffic volume and percentage of traffic reduction compared to granularity 1, for different granularity in PageRank.}
    \label{Tab:pagerank-traffic}
    \begin{tabular}{l|ccccccc}
        \toprule
        \textit{Granularity} & \textbf{1} & \textbf{2} & \textbf{4} & \textbf{8} & \textbf{16} & \textbf{32} & \textbf{64} \\
        \midrule
        \textit{Traffic (GiB)} & 3068 & 1532 & 764 & 380 & 188 & 92 & 44 \\
        \textit{\% Reduction} & n/a & 50.0\% & 75.0\% & 87.6\% & 93.8\% & 97.0\% & 98.5\% \\
        \bottomrule
    \end{tabular}
\end{table}

\subsubsection{TeraSort}

We have implemented TeraSort based on the MapReduce model from the Hi-Bench suite~\cite{huang_hibench_2010}.
TeraSort is of particular interest because it involves a heavy data shuffle phase.
We want to compare a version of TeraSort following the serverless MapReduce approach~\cite{pu_shuffling_2019,sanchez_primula_2020,aws_lambdamapreduce_2018}, with a single-stage burst computing version, where we take advantage of locality for the shuffle phase.
The main advantages of burst are:
\begin{inparaenum}[(i)]
    \item the MapReduce version requires two rounds of function invocations (map and reduce), while the burst model requires a single flare, and
    \item the MapReduce version shuffles data through object storage, while burst employs the (locality-aware) \textit{all-to-all} collective.
\end{inparaenum}

\paragraph{Setup}
This experiment sorts a \gbyte{100} dataset generated with Hi-Bench~\cite{huang_hibench_2010} with 192 partitions.
The burst platform runs on EKS with two \texttt{m7i.48xlarge} (96 vCPUs, 384 GB) invokers and an \texttt{c7i.xlarge} controller.
The input data is in an Amazon S3 bucket located in the same region.

\begin{figure}
    \centering
    \begin{subfigure}{\linewidth}
        \includegraphics[]{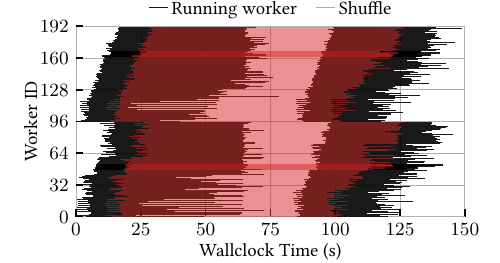}
        \caption{MapReduce}
        \label{Fig:terasort-mr}
    \end{subfigure}
    \begin{subfigure}{\linewidth}
        \includegraphics[]{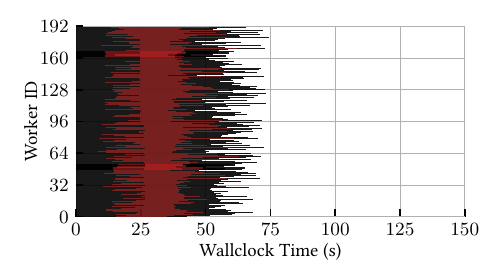}
        \caption{Burst}
        \label{Fig:terasort-burst}
    \end{subfigure}
    \caption{
    TeraSort timeline comparison between \ref{Fig:terasort-mr} serverless MapReduce and \ref{Fig:terasort-burst} burst computing.
    MapReduce comprises two function rounds (map and reduce), with data exchange via object storage.
    Burst uses a single flare, exchanging data through the \textit{all-to-all} collective.
    }
    \label{Fig:terasort}
\end{figure}

\paragraph{Results}
\cref{Fig:terasort} shows the timeline of two executions comparing both models.
The execution time of each worker is shown in horizontal black bars, stacked by \textit{Worker ID} on the vertical axis.
Superimposed in red, we see the time elapsed for the shuffle phase of the TeraSort algorithm.
In the MapReduce version (\cref{Fig:terasort-mr}), we highlight
\begin{inparaenum}[(i)]
    \item the dispersity in function start-up time, as we have seen in \cref{Fig:invocation-simult};
    \item a gap where no functions are running, caused by splitting the workload into two phases (map and reduce), with an externally-managed synchronization phase between them, adding further overhead; and
    \item an outlier in the map phase (worker $\#121$), which slows down the entire workflow.
\end{inparaenum}

All these points are addressed in burst computing (\cref{Fig:terasort-burst}).
First, the group invocation packs all workers into two containers of 96 workers, making start-up faster, and ensuring parallelism, which eliminates (latency-induced) outliers.
Second, worker-to-worker collectives avoid splitting the workload into two phases.
Finally, remote communication is reduced as shown in \cref{Sec:experiment_group_collectives} thanks to locality.
To wit, we achieve a final speed-up of $~2\times$ for this particular execution, but we attained a mean speed-up of $1.91\times$ across six executions.

\paragraph{Takeaway}
Frictions \textbf{F1}, \textbf{F2}, and \textbf{F3} appear in real-world applications.
Hyperparamenter tuning shows duplication in worker initialization due to friction \textbf{F1}, which also slows down worker start-up in PageRank and TeraSort.
PageRank and TeraSort evidence the issues of friction \textbf{F2}: the iterative nature of PageRank makes it unfeasible in FaaS due to excessive stages, and TeraSort is hindered considerably due to slower coordination.
Burst computing mitigates this by allowing workers to coordinate and share data in a single stage, instead of requiring multiple stages orchestrated externally that communicate asynchronously through storage.
PageRank and TeraSort also emphasize friction \textbf{F3}: Page-Rank with iterative, large communication to aggregate the vector and TeraSort with a single, large data shuffle.

\section{Related work}
\label{RelatedWork}

As introduced in \cref{Introduction}, the concept of resource burstability has already been discussed in the serverless literature.
FaaS has been referred to as a ``burstable supercomputer-on-demand''~\cite{fouladi_encoding_2017} for ``burst-parallel serverless applications''~\cite{thomas_particle_2020}, albeit with limitations.
``Flash bursts'' in HPC~\cite{li_millisort_2021} explore the feasibility of \millisecond{1} jobs spanning a large number of servers, which is attractive but unfeasible with existing technologies in public clouds.
The authors use a highly-optimized dedicated cluster with low-latency communication and they neglect issues such as application loading, isolation, and multi-tenancy, among others, which are key in, e.g., FaaS.
\citet{muller_serverless_2020} pursue serverless burstability for batch jobs and envision Serverless Clusters, highlighting that functions are too limited for the job.

Burst computing goes a step further in defining a new way of running burst-parallel jobs in the cloud.
It evolves from FaaS to exploit serverless burstiness but it unlocks the limitations of working with functions to provide a compute environment tailored for massively parallel collaborative jobs.
To the best of our knowledge, we are the first ones to raise the unit of management from a function to the job level in a serverless service with a group invocation abstraction that allows to pack workers together and enable locality within the job.

Several papers have explored how to reduce function start-up time in FaaS.
For example, SAND~\cite{akkus_sand_2018} and Faastlane~\cite{kotni_faastlane_2021} use a single container for all function invocations in the same application.
However, they become limited in scalability and unfit for burst workloads that require to distribute workers among different machines.

MxFaaS~\cite{stojkovic_mxfaas_2023} aims at optimizing a FaaS runtime for burst jobs in a transparent way.
Its MxContainer improves performance by sharing processor cycles, I/O bandwidth and memory state between invocations of the same function.
%
This shows that burst jobs are becoming widely accepted in FaaS settings, with locality as a key performance enhancement.

Other recent works aim to pack multiple function invocations into the same container to improve invocation latency and resource usage~\cite{basuroy_propack_2023, wu_faasbatch_2024, jin_ditto_2023}.
However, all these solutions (including SAND, MxFaaS, etc.) still lack the concept of group invocations, they run functions as individual invocations, and thus fail to provide optimizations in resource provisioning and parallel guarantees.

Communication and state sharing in FaaS has been widely explored in the literature.
Some works like Cloudburst~\cite{sreekanti_cloudburst_2020} or Faasm~\cite{shillaker_faasm_2020} reimagine a FaaS platform to integrate a shared data space where multiple functions may put their common state.
Others solve this with dedicated remote storage solutions, e.g., Pocket~\cite{klimovic_pocket_2018}, Crucial~\cite{barcelonapons_crucial_2019}, or Glider~\cite{barcelonapons_glider_2023}.
These solutions are orthogonal to burst computing.

Boxer~\cite{wawrzoniak_boxer_2021} explores direct communication, and FMI~\cite{copik_fmi_2023} builds a library of collectives between groups of FaaS functions (with NAT traversal).
They only tackle current FaaS platforms and do not provide any locality optimizations, as possible in burst computing and exploited by our BCM.
FMI could be used in burst computing as a remote backend in the BCM to accelerate pack-to-pack transfers.

Finally, we want to outline a clear trend in FaaS that is aligned with burst computing: the emergence of Big Functions (Big Lambdas) with more than one CPU core (currently up to 6 vCPUs in AWS).
With these bigger functions, jobs may also leverage function intra-parallelism to scale computations more efficiently.
Similar to burst computing, worker locality becomes relevant and improves the execution of recursive data processing algorithms that use aggregations or hierarchical approaches to problem parallelization and communication.
However, Big Functions are still FaaS and lack group-aware mechanisms and parallel guarantees.

\section{Conclusion}
\label{Conclusion}

In this paper, we have presented burst computing, a novel cloud computing model designed to address the growing demand to run sudden, variable, burst-parallel workloads without provisioning resources in advance.
We have reviewed the challenges and shortcomings of current technologies (i.e., FaaS), and we have demonstrated the effectiveness and versatility of our proposed solution.

Burst computing offers several key advantages over existing FaaS technologies, primarily the addition of a group invocation primitive that allows the platform to manage jobs as a unit, instead of independent function invocations.
This raises tenant isolation to the job level and allows to allocate resources en masse and apply worker packing, which in turn enables powerful locality between workers.
Our experiments and performance evaluations have shown that our platform achieves significant improvements by exploiting this locality in job invocation latency, worker simultaneity, code and data loading, and worker-to-worker communication with group collectives.
Further, we demonstrated speed-ups of $13\times$, and $2\times$ in PageRank, and TeraSort, respectively, thereby validating efficacy in real-world scenarios.

In conclusion, burst computing represents a significant advancement for serverless data processing in the cloud, with potential to become a new paradigm of cloud services.
It unlocks key limitations of FaaS, becoming the next step forward to support applications previously stymied by its restrictive model~\cite{hellerstein_serverless_2018}.
We believe that our contributions are just the substrate for further innovation such as new workflow definition tools and orchestration engines that leverage burst computing jobs and strive towards a simplified dynamic utilization of cloud resources.

\begin{acks}
    This work is partially funded by the \grantsponsor{europe}{Horizon Europe programme}{} under grant agreements No.~\grantnum{europe}{101092644} (Neardata), \grantnum{europe}{101092646} (CloudSkin), and \grantnum{europe}{101093110} (Extract);
    and by the \grantsponsor{spain}{Spanish Ministry of Economic Affairs and Digital Transformation}{} and the European Union-NextGenerationEU (within the PRTR and MRR framework) through the \grantnum{spain}{CLOUDLESS UNICO I+D CLOUD 2022}.
\end{acks}

\bibliographystyle{ACM-Reference-Format}
\bibliography{Burst}



\end{document}